\documentclass[11pt]{article} 

\usepackage[a4paper, margin=1in]{geometry} 

\usepackage{graphicx}%
\usepackage{multirow}%
\usepackage{amsmath,amssymb,amsfonts}%
\usepackage{amsthm}%
\usepackage{mathrsfs}%
\usepackage{comment}%
\usepackage[title]{appendix}%
\usepackage{xcolor}%
\usepackage{textcomp}%
\usepackage{manyfoot}%
\usepackage{booktabs}%
\usepackage{algorithm}%
\usepackage{algorithmicx}%
\usepackage{algpseudocode}%
\usepackage{listings}%
\usepackage{makecell}
\usepackage{titlesec}            
\usepackage{authblk}             
\usepackage{hyperref} 

\theoremstyle{thmstyleone}%
\theoremstyle{thmstyletwo}%

\theoremstyle{thmstylethree}%
\date{\vspace{-4.0em}}
\def\etal{\emph{et al. }}

\begin{document}

\def\fnm#1{\leavevmode\hbox{#1}}%
\def\sur#1{\unskip~\nobreak\leavevmode\hbox{#1}}%
\def\spfx#1{#1}%
\def\pfx#1{#1}%
\def\sfx#1{#1}%
\def\tanm#1{#1}%
\def\dgr#1{#1}%

\title{Towards a Generalizable Speech Marker for Parkinson's Disease Diagnosis}

\author[1,2]{Maksim Siniukov\thanks{\href{mailto:siniukov@usc.edu}{siniukov@usc.edu}}}
\author[1,2]{Ellie Xing\thanks{\href{mailto:exing@ict.usc.edu}{exing@ict.usc.edu}}}
\author[3]{Sanaz Attaripour Isfahani\thanks{\href{mailto:sattarip@hs.uci.edu}{sattarip@hs.uci.edu}}}
\author[1,2]{Mohammad Soleymani\thanks{\href{mailto:soleymani@ict.usc.edu}{soleymani@ict.usc.edu}}}

\affil[1]{Institute for Creative Technologies, University of Southern California, Playa Vista, CA 90094, USA}
\affil[2]{Thomas Lord Department of Computer Science, University of Southern California, Los Angeles, CA 90089, USA}
\affil[3]{School of Medicine, University of California Irvine, Irvine, CA 92697, USA}
\maketitle
\abstract{Parkinson’s Disease (PD) is a neurodegenerative disorder characterized by motor symptoms, including altered voice production in the early stages. Early diagnosis is crucial not only to improve PD patients' quality of life but also to enhance the efficacy of potential disease-modifying therapies during early neurodegeneration, a window often missed by current diagnostic tools. In this paper, we propose a more generalizable approach to PD recognition through domain adaptation and self-supervised learning. We demonstrate the generalization capabilities of the proposed approach across diverse datasets in different languages. Our approach leverages HuBERT, a large deep neural network originally trained for speech recognition and further trains it on unlabeled speech data from a population that is similar to the target group, i.e., the elderly, in a self-supervised manner. The model is then fine-tuned and adapted for use across different datasets in multiple languages, including English, Italian, and Spanish.
Evaluations on four publicly available PD datasets demonstrate the model’s efficacy, achieving an average specificity of 92.1\% and an average sensitivity of 91.2\%. This method offers objective and consistent evaluations across large populations, addressing the variability inherent in human assessments and providing a non-invasive, cost-effective and accessible diagnostic option.}%

\vspace{0.5cm}

\section{Introduction}\label{sec1}

\label{sec:intro}

Parkinson's Disease (PD) is the most common neurodegenerative disease after Alzheimer's, affecting over 10 million people worldwide~\cite{standaert2015parkinson}. PD results from the progressive loss of neurons in various brain regions, most notably the substantia nigra. This degeneration results in alterations in the levels of neurotransmitters~\cite{birkmayer19613, kakkar2015management}. The primary neurotransmitter affected in PD is dopamine due to the loss of dopaminergic neurons. Alterations in acetylcholine, serotonin, and noradrenaline also occur~\cite{bosboom2003role,buddhala2015dopaminergic,delaville2011noradrenaline, fornai2007noradrenaline}. Lack of dopamine disrupts the balance of the neural circuits that control movement, leading to symptoms such as bradykinesia, rigidity, and tremor.

Although PD remains incurable, early diagnosis followed by treatment can improve the quality of life of patients with PD~\cite{braga2019automatic}. Early diagnosis is crucial because it allows for interventions when neurons in the substantia nigra are not yet fully degenerated. This stage increases the potential effectiveness of emerging therapies that aim to stop or reverse the disease process, a significant focus of current research. 

PD is diagnosed primarily through clinical findings, supplemented by various non-invasive and invasive methods. These methods include Dopamine Transporter Single-Photon Emission Computed Tomography (DAT-SPECT), cerebrospinal fluid analysis, and skin biopsy, which may enhance diagnostic precision. However, the definitive gold standard for PD diagnosis remains post-mortem examination (autopsy). Many of the currently available diagnostic tools, including some imaging and fluid biomarkers, are still being studied and are not yet fully established for routine clinical use~\cite{coughlin2023fluid, donadio2020skin}.

There are also non-invasive methods that can reduce the cost and enhance the clinical accuracy of PD detection. These methods, such as voice analysis and wearable sensing, are simpler to apply, low-cost, and allow for remote diagnosis and health monitoring~\cite{little2008suitability}. Vocal disorders are common in PD patients and can be early indicators of the disease, alongside other motor symptoms~\cite{rusz2015speech}; this has motivated research on voice-based Parkinson's diagnosis~\cite{ibarra2023towards, Reddy}.

Voice alterations resulting from PD include hypokinetic dysarthria that may be present as irregular pauses, dysphonia, reduced fundamental frequency range, hypophonia (reduced vocal volume), monotone (reduced variability of pitch), dysarthria (difficulty with articulation), tremor and dyskinesia (including uncontrolled lip movements), bradykinesia (slowness of speech), and rigidity (muscle stiffness)~\cite{sakar2013collection,berardelli2001pathophysiology,jankovic2005motor,shahed2007motor}. Therefore, speech processing is a promising approach for early PD diagnosis and monitoring symptom progression for treatment adjustment~\cite{Amato2023}. Conditions commonly comorbid with PD, such as fatigue, anxiety and depression as well as other PD-related symptoms like tremors, can also affect speech and should be accounted for in voice analysis.

Existing work on machine learning for the diagnosis of PD relies on datasets collected with controlled vocalization of sustained vowels, numbers, single words and short sentences in addition to phrases with varying degrees of articulation difficulty~\cite{Spanish,ozbolt2022things,Italian, ItalianPaper, orozco2014new, London, bot2016mpower}. Voice descriptors for capturing articulatory and phonatory aspects of speech have been successfully used to recognize PD from healthy control (HC) in previous studies, including jitter~\cite{quan2022end}, shimmer, instantaneous energy deviation cepstral coefficient~\cite{karan2020hilbert}, Hilbert cepstral coefficients~\cite{karan2021improved}, and Mel Frequency Spectral Coefficients (MFCC)~\cite{hawi2022automatic, lahmiri2018performance} around voiced and unvoiced segment boundaries. 
Various machine learning models are used for PD diagnosis including ensemble-based methods, support vector machines~\cite{shahbakhi2014speech_geneticsvm,nonavinakere2021detection,tsanas2012novel,sakar2013collection,lahmiri2018performance,zhang2021parkinson}, naive Bayes~\cite{lahmiri2018performance}, sparse representation~\cite{Reddy} and k-nearest neighbors~\cite{sakar2013collection, tuncer2020automated}.

Earlier approaches to automatic PD detection from speech used classifiers trained with a set of chosen hand-crafted speech features, focusing on articulation and phonation aspects of speech. Phonation features often include jitter, shimmer, harmonic-to-noise ratio (HNR), noise-to-harmonic ratio (NHR), pitch and Mel-frequency cepstral coefficients (MFCCs)~\cite{Velazquez}. For instance, a support vector machine (SVM) classifier trained with traditional phonation measures and pitch period entropy effectively distinguished PD from healthy subjects based on sustained phonations~\cite{Little}. 

Recent work has used deep learning models for identifying speech biomarkers of PD. Deep learning models are able to automatically learn speech descriptors from large and varied datasets, making them effective across different languages and demographics. In particular, recent end-to-end models are able to accept digitized audio with little or no transformation. Different deep learning models have shown their high capabilities for speech-based Parkinson's detection including Long-Short-Term Memory neural networks (LSTM)~\cite{fang2020parkinsonian,quan2021deep,ibarra2023towards,er2021parkinson}, recurrent neural networks (RNN)~\cite{vasquez2018multitask}, transformers~\cite{klempivr2023evaluating, klempir2024analyzing, escobar2023deep}, convolutional neural networks (CNN) ~\cite{gunduz2019deep,wodzinski2019deep,ibarra2023towards,er2021parkinson,escobar2023deep,arias2020predicting}, multi-layer perceptions (MLP)~\cite{sainath2015learning,parisi2018feature, arias2020predicting}. 
Existing work~\cite{wodzinski2019deep,vasquez2019convolutional, arias2020predicting, rios2020transfer, vasquez2021transfer, escobar2023deep, ibarra2023towards, klempivr2023evaluating, klempir2024analyzing} commonly use transfer learning to adapt knowledge from models trained on a large dataset on a relevant task, e.g., speech recognition.

Factors such as language, age and cognitive impairment might affect the distribution of speech features and, as a result, the generalizability of speech-based PD recognition~\cite{AriasVergara2017,Garcia2021}. Arias-Vergara \etal~\cite{AriasVergara2017} found that phonation and articulation capabilities differ between younger and older subjects, regardless of PD status, thus necessitating age matching or adaptation for PD recognition models. 

Ibarra \etal~\cite{ibarra2023towards} used end-to-end deep learning for PD detection from speech. Their approach proved to be able to detect PD across different datasets, highlighting such models' potential to generalize beyond a single dataset. Using diverse datasets for training helps models to be flexible and consider different demographics, languages, and recording conditions, as machine learning models can suffer from performance degradation on different data or populations due to shifts in data distribution. The development of corpus-independent models requires understanding how sensitive speech features are to such shifts between different languages or instrumentation. The implementation of transfer learning techniques offers a potential solution by fine-tuning an existing model trained on large and diverse data. Such a general approach helps PD detection models adjust better for diverse recordings and enhances the model's robustness and applicability in real-world scenarios. 

We present a novel approach to solve these challenges. First, we adapt HuBERT~\cite{hubert}, a powerful large pre-trained speech model to the elderly's speech. We perform this adaptation alignment via domain adaptive pretraining (DAPT)~\cite{dontstoppretrain} on a large corpus of recordings of elderly people. Second, we apply domain adversarial training (DAT)~\cite{ganin2015unsupervised} to improve generalizability across languages by learning language-agnostic features. Our method achieves state-of-the-art performance for speech-based PD detection in several languages and presents a significant step towards advancing speech-based PD screening. 

\section{Materials and Methods}
\subsection{Datasets}
We leverage two sets of data for training and evaluating our models. First, unlabeled datasets that we use for adapting general speech models to the target population, including VoxCeleb~\cite{voxceleb} and mPower~\cite{bot2016mpower}. 
Second, datasets with clinically valid PD diagnosis. The demographic information of the datasets with PD labels is summarized in Table~\ref{tab:datasets}. Below, you will find a detailed description of the datasets.

     \textbf{PD-Neurovoz dataset.} The Neurovoz dataset~\cite{Spanish} contains Spanish speech with Parkinson's Disease diagnosis. There are samples of hypokinetic dysarthria, which highlights PD's impact on articulation. The Neurovoz dataset has recordings from 47 Parkinsonian patients and 32 healthy control speakers, all native speakers of Castilian Spanish. The data was gathered at the otorhinolaryngology and neurology services of the Gregorio Marañón Hospital in Madrid, Spain. The speech task was to repeat a syllable sequence~\cite{Spanish,ozbolt2022things}.
    Unified Parkinson’s Disease. Rating Scale (UPDRS) and Hoehn \& Yahr (H\&Y) scales were used as the diagnosis tool.
    
     \textbf{PD-Italian dataset.} The PD-Italian dataset~\cite{Italian, ItalianPaper} comprises recordings from three groups: Young Healthy Controls (YHC), Healthy Elderly Controls (HEC), and PD patients. From the YHC group, there were 15 participants aged 19 to 29 years old, mostly male. The HEC group consisted of 22 participants from 60 to 77 years old. The data contains 28 PD patients between 40 and 80 years old. Participants were instructed to read a text and repeat syllables. UPDRS diagnosis tool was used as in PD-Italian dataset.
    
     \textbf{PD-GITA dataset.} PD-GITA dataset~\cite{orozco2014new} was collected through recordings of spontaneous speech from a total of 110 Spanish speakers from Colombia, including 50 healthy control subjects, 25 Depressive Parkinson's Disease (D-PD) patients, and 35 Non-Depressive Parkinson's Disease (ND-PD) patients. All patients were diagnosed using UPDRS and H\&Y scales. Participants were asked to talk about their daily routines, which is the same task used to evaluate depression in the Movement Disorders Society – Unified Parkinson's Disease Rating Scale (MDS-UPDRS). The average duration of the monologues is 84 $\pm$ 34 s for D-PD patients, 80 $\pm$ 37 s for ND-PD patients, and 45 $\pm$ 24 s for health subjects, totaling 2 hours and 3 minutes of recordings. Transcriptions were manually created following the verbatim protocol, with the total vocabulary of the transcriptions, excluding stop-words, containing 1418 words.
    
     \textbf{MDVR-KCL.} The Mobile Device Voice Recordings at King's College London (MDVR-KCL) dataset~\cite{London} has a collection of voice recordings from individuals diagnosed with early and advanced Parkinson's disease as well as healthy controls. The participants were given 2 tasks: to read “The North Wind and the Sun” and “Tech. Engin. Computer applications in geography snippet" paragraphs of text and to have a spontaneous dialog with the doctor. Recordings are labeled with metadata, such as health status, H\&Y scale rating, and scores from UPDRS scales.~\cite{orozco2014new}

     \textbf{mPower dataset.} Data collection for mPower database~\cite{bot2016mpower} involved using a specially designed mPower mobile app; the recorded voice involved participants sustaining the sound ‘Aaaaah’ into the microphone at a constant volume for a duration of up to 10 seconds. The data from this task contains audio files with measurements taken from the iPhone microphone. A total of 9,520 individuals consented to participate in the study and agreed to share their data with the scientific community broadly. Out of the 6,805 participants who filled out the enrollment survey, 1,087 reported clinical diagnoses of PD, while 5,581 did not (137 chose not to respond to the question). This led to a total of 65,022 recordings, amounting to 168 hours of speech data. 
     
     \textbf{VoxCeleb2 dataset.} 
    VoxCeleb2 dataset~\cite{chung2018voxceleb2} is a large audiovisual dataset originally designed for speaker recognition. The dataset consists of 150,480 YouTube videos featuring a total of 6,112 speakers. The compilation process involved a multi-stage approach: actively verifying the speaker through a two-stream synchronization Convolutional Neural Network (CNN), followed by confirming the speaker's identity with a CNN-based facial recognition system. This methodology ensures high-quality data and accurate labeling.
    
     \textbf{AgeVoxCeleb dataset.}
     Age-VOX-Celeb dataset~\cite{tawara2021agevox} builds upon VoxCeleb2 video sequences~\cite{chung2018voxceleb2} and estimates people's age by combining meta-data from annotations and video description, wrong age label cleaning by facial age estimation and manual cleaning.
    We use age information from Age-VOX-Celeb to identify the speakers who are over 60 years old in VoxCeleb2 for pertaining.

\begin{table}[!t] %
\centering %
\begin{minipage}{0.88\textwidth}
\caption{Dataset information, including the number of participants and mean age, is presented for the PD-GITA, PD-Neurovoz, PD-Italian, MDVR-KCL, and mPower datasets. For the PD-GITA, PD-Neurovoz, PD-Italian, and MDVR-KCL datasets. Patients with Parkinson's Disease are denoted as PD; healthy controls are denoted as HC. For the mPower dataset, diagnosis-specific statistics are unavailable; therefore, the table provides the total number of subjects and their average age without a distinction between PD and HC. \label{tab:datasets}}%
\vspace{1.0em}
\begin{tabular*}{\textwidth}{@{\extracolsep\fill}lcccccc@{\extracolsep\fill}}%
\toprule
\textbf{Corpus} & \multicolumn{2}{c}{\textbf{\makecell{Subjects\\ (PD/HC)}}} & \multicolumn{2}{c}{\textbf{\makecell{Age\\ (PD/HC)}}} & \textbf{\makecell{Total\\duration\\(min)}} & \textbf{\makecell{Average\\segment\\ duration}} \\
 & \textbf{F} & \textbf{M} & \textbf{F} & \textbf{M} &  & \textbf{(sec)} \\
\midrule
PD-GITA & 25/25 & 25/25 & 60/61 & 62/61 & 230.6 & 3.2 \\
PD-Neurovoz & 21/23 & 24/24 & 70/70 & 67/61 & 215.0 & 4.4 \\
PD-Italian & 41/44 & 47/44 & 67/63 & 67/64 & 254.4 & 19.4 \\
MDVR-KCL & 4/18 & 12/3 & -/- & -/- & 114.0 & 93.7 \\
mPower & 1461 & 5314 & 39 & 36 & 11753.4 & 12.9 \\ 
\bottomrule
\end{tabular*}
\end{minipage}
\end{table}

\subsection{Method}

\begin{figure}[t]
\centering
\begin{minipage}{0.88\textwidth}
\includegraphics[trim=0 140 0 62, clip, page=1, width=1.\textwidth]{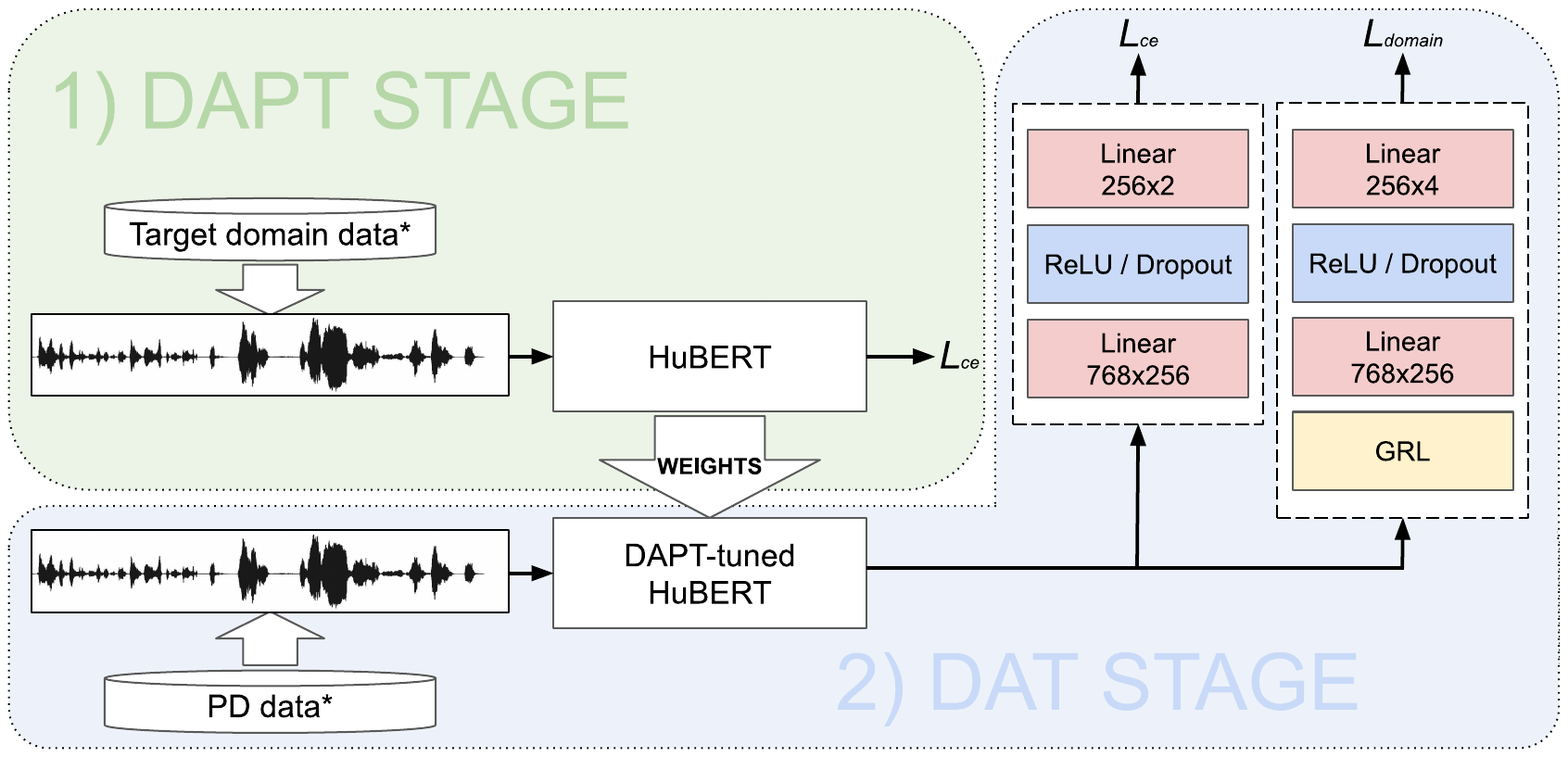}
\caption{Overview of the pipeline. In the first stage, HuBERT Domain Adaptive Pretraining (DAPT) is performed on the HuBERT model to adapt to the target domain data containing speech from the elderly$^{*}$. Target domain data may be either mPower dataset or AgeVox dataset. The learning objective for DAPT is masked-units cross-entropy loss. DAPT-tuned weights are transferred to the second stage. In the second stage, Domain Adversarial Training (DAT) is applied to PD datasets$^{**}$. DAT objectives are PD prediction cross-entropy loss and domain discrimination cross-entropy loss. A Gradient Reversal Layer (GRL) is placed between the domain classifier and HuBERT networks. \\$^{*}$ Target domain data may be: mPower dataset or AgeVox dataset.
$^{**}$ PD datasets are: PD-GITA, PD-Neurovoz, PD-Italian, and MDVR-KCL.} 
\label{fig:model}
\end{minipage}

\end{figure}
In this section, we provide an overview of our pipeline. We first describe HuBERT~\cite{hubert}, a powerful pre-trained speech encoder that serves as a foundational component of our model. Second, we describe domain adaptive pretraining -- a technique that adapts the pre-trained speech model with elderly demographics of PD datasets. Finally, we leverage domain adversarial training to bridge the gap between languages to make the model focus on language-agnostic features, which makes the method more generalizable to unseen/novel speakers. Our pipeline is shown in Figure~\ref{fig:model} 

\subsubsection{Pre-trained speech encoder}
In this work, we choose to use HuBERT~\cite{hubert}, a powerful pre-trained speech encoder inspired by BERT, a popular model that was developed for language modeling. HuBERT was originally developed for speech recognition but is able to achieve high performance in various speech understanding tasks, including automatic speech recognition, language generation, and speech compression. 

HuBERT pretraining consists of two stages of representation learning. In the first stage, pseudo-labels are generated via applying K-means clustering with 100 clusters on 39-dimensional MFCC features~\cite{mfcc_davis1980comparison}, and the HuBERT model learns to predict those cluster indices (pseudo-labels) given partly masked speech frames (tokens). In the second stage, k-means clustering with 500 clusters was performed on pre-trained stage-one HuBERT Large latent features to obtain better pseudo labels. Then, the HuBERT base is learned to predict those cluster classes in a self-supervised way. 
    Data used for pretraining on both stages was 960-hour LibriSpeech.
\subsubsection{Domain adaptive pre-training}
Original HuBERT was trained on LibriSpeech~\cite{panayotov2015librispeech} and Libri-light~\cite{kahn2020libri} datasets, which contains read-aloud speech from English audiobooks. Those datasets contain recordings of people of different ages, with little representation of the elderly. In contrast, Parkinson's disease generally affects older people, especially people who are over 60 years old.~\cite{willis2013parkinson}. In addition, a PD diagnosis tool should be able to work in different languages, not only English. Both factors result in a large domain gap between two datasets: LibriSpeech -- dataset that the original HuBERT was trained on, and the PD datasets. To address this problem, we perform HuBERT Domain adaptive pretraining following \cite{dontstoppretrain}. 
First, we utilize the Age-VOX-Celeb dataset~\cite{tawara2021agevox} -- dataset extracted from VoxCeleb2 video sequences~\cite{chung2018voxceleb2}  -- by taking only sequences with people who are older than 60 years old to create elderly people VoxCeleb 2 subset. The VoxCeleb dataset is multilingual, with speakers of 145 different nationalities~\cite{voxceleb}, which ensures that the domain-adapted HuBERT model acquires features that are not just pertinent to elderly people but are also linguistically diverse. The resulting subset contains 16\% of the Age-VOX-Celeb dataset that contains 168k audio clips, making it suitable for model alignment to the cross-lingual PD detection domain.
 
We continue pretraining the HuBERT base from the original checkpoint following the original paper's pipeline. First, we extract features from the Age-VOX-Celeb subset utilizing the HuBERT large model. Subsequently, these features are processed through a pretrained k-means clustering model to generate pseudo-labels. These pseudo-labels serve for domain-adaptive pretraining~\cite{dontstoppretrain}, where we employ a cross-entropy loss calculated over the pseudo-labeled segments that have been selectively masked. 

We denote the utterance of speech frames as \( X \), where \( X \) has a length of \( T \). Let \( M \) represent the masked token indices. The corrupted version of \( X \) is denoted as \( \hat{X} \), defined as:

\[
\hat{X}_i = 
\begin{cases} 
X_i, & \text{if } i \notin M \\
\hat{x}, & \text{if } i \in M
\end{cases}, \quad \forall i \in \{1, \dots, T\}
\]

Specifically, the masked frames are replaced with a mask embedding vector \( \hat{x} \). The HuBERT model predicts a probability distribution \( P_H \), which represents the probability that the utterance corresponds to each cluster, including the correct cluster \( z_t \). The model is trained using the masked cross-entropy loss function, defined as:

\[
\mathcal{L}_{\text{DAPT}} = -\sum_{t \in M} \log P_H(z_t \mid \hat{X})
\]

This enables the model to adjust to the domain-specific characteristics of speech in the elderly population, enhancing its ability to learn features that are more relevant to PD detection in the target domain.

\subsubsection{Baseline PD network}
After domain adaptive pretraining, HuBERT serves as a powerful feature extractor. A multilayer perceptron (MLP) for sequence classification is put on top of mean-pooled HuBERT encoder features. 
The whole model is fine-tuned for PD detection on the labeled datasets using binary cross-entropy loss. This architecture serves as a baseline and is further extended in domain adversarial training.
\subsubsection{Domain adversarial training}
We use a DAT domain adaptation method to adapt the pretrained HuBERT model to the target data. Domain adaptation techniques change the model weights in a way that improves its performance on target data, in this case, an unlabeled database with PD and healthy control. Specifically, we use Domain-Adversarial Neural Network (DANN)~\cite{ganin2015unsupervised} that consists of four parts: joint feature extractor, PD classifier, gradient reversal layer (GRL), and domain classifier. The gradient reversal layer and the domain classifier force the model to learn features invariant to the dataset, i.e., source training data and the target unseen and unlabeled test data.
We use the same feature extractor and PD classifier as in the baseline PD network (DAPT pretrained HuBERT) while adding a domain classifier on top of HuBERT to determine which dataset the output of the encoder is from. 
The goal of DANN is to minimize classification loss for the PD detection task while maximizing domain classifier loss, making features invariant to domains. In this way, the encoder (deep feature extractor) learns features that are useful for PD detection and do not distinguish between different languages, microphones or groups of people.

\subsubsection{Baselines}
We compare our model against three baseline methods: traditional machine learning (ML) pipelines with hand-crafted features, exemplar-based sparse representation classification, and a 1D convolutional neural network (CNN).

We use two popular ML classifiers known for their efficacy in handling complex and noisy data: support vector machine (SVM), a supervised learning algorithm that classifies data points by maximizing the margin between the samples from different classes in a high-dimensional space, and Random Forest, an ensemble learning method that aggregates multiple decision trees to output the mode of the classes.

Exemplar-based sparse representation (SR) classification, introduced by Reddy et al.,\cite{Reddy}, involves representing test data using a sparse combination of exemplars (speech feature vectors) extracted from the training data, bypassing traditional ML training. SR classification based on l1-regularized least squares (LSRC) utilizes a dictionary containing data from both classes (PD and HC). Reddy \etal LSRC~\cite{Reddy} employs multiple class-specific dictionaries. Solving the L1-regularized least squares problem with the dictionary or dictionaries derives the sparse coefficient vector, which is then interpreted sparsely.

The 1D CNN architecture, proposed by Ibarra \etal~\cite{ibarra2023towards}, utilizes multiple 1D CNN layers and a convolutional attention mechanism to capture temporal patterns in Mel-scale spectrograms extracted from 40ms speech segments. It is an effective use of deep learning to capture temporal patterns in speech and serves as a good baseline model for our task.

Since there was no baseline source code for LSRC, Reddy \etal's LSRC~\cite{Reddy}, Random Forst or SVM methods we re-implemented those methods based on the description from the paper. In order to be as close to the original description as possible, we trained and evaluated separate models for each dataset independently instead of having one model for all datasets. In the case of the 1CNN model, we used the one-for-all-domains model in the same way as we evaluated our model.

\subsection{Data preprocessing}
Datasets for Parkinson's disease detection contain a significant amount of noise: The MDVR-KCL dataset includes segments of a doctor conversing over the phone with a patient, as well as the phone's ringing, which is irrelevant to PD detection. We purged the dataset of such occurrences.
All datasets contain long segments of silence, which are not relevant for Parkinson's disease detection. We employed an energy-based method to remove these silent segments.
To this end, first, all the recordings are resampled to 16 kHz. Then, the energy level of the sound over an interval is calculated using the root mean square (RMS) method, where the RMS value is computed for all the signal amplitudes within a window of 481 samples. This window size was found optimal for a 16 kHz sampling rate. To determine if a segment of the audio is silent, we compare the RMS energy level against a predefined threshold. In our method, the threshold is set to 0.0025.

Not all silent segments are removed, as some may be brief pauses in speech that define the pace of a person and could be valuable for Parkinson's detection. Hence, we only eliminate segments that exceed 500 milliseconds in duration.

\subsection{Training and evaluation Details}
During domain adversarial training, the model training incorporated LayerDrop set at a rate of 0.1. The AdamW optimizer was utilized, with a learning rate of $3 \cdot 10^{-5}$, and a linear learning rate scheduler. Model training was conducted for 80 epochs with a batch size of 128 samples. To mitigate the effects of excessive gradient magnitudes, gradient clipping was applied with a maximum gradient norm of 1. The training was performed on a single NVIDIA A40 GPU. 
The gradient reversal layer $GRL$ proposed by Ganin \etal~\cite{ganin2015unsupervised} is placed between the domain classifier $E_{D}$ and HuBERT encoder $E$. The GRL layer reverses the domain classification loss gradient direction and scales the gradient values by a positive factor $\lambda$ when it passes through GRL layer, which gives control over the balance between the domain classifier and the encoder objectives.

\[
\frac{\partial E(x)}{\partial \theta_E} =   -\frac{\partial \text{GRL}(E(x))}{\partial \theta_E} 
\]
where $x_i$ is a speech utterance, $\theta_E$ is HuBERT encoder parameters. 
This way, the HuBERT encoder learns to maximize the domain classification error, and the domain classifier learns to minimize it.

\[
\mathcal{L}_{\text{Domain}} = -\sum_{i=1}^N \log P(d_i \mid E_{D}(\text{GRL}(E(x_i))))
\]

\[
\mathcal{L}_{\text{PD}} = - \sum_{i=1}^N \log P(y_i \mid E_{PD}(E(x_i)))
\]

\[
\mathcal{L}_{\text{DAT}} = \mathcal{L}_{\text{PD}} + \lambda \cdot \mathcal{L}_{\text{Domain}}
\]
where $x_i$ is speech utterance, $y_i$ is corresponding PD label, and $d_i$ is corresponding domain label.

The optimal value of GRL $\lambda$ was found to be 0.1 via grid search on \{0.01, 0.05, 0.1, 0.5, 1, 2\} grid; we use the optimal value in all of the experiments. 

For Domain adversarial training experiments, $E_{PD}$ and $E_D$ are MLP-based. Each encoder begins with a fully connected layer that maps inputs of size 768 to outputs of size 256, followed by mean-pooling along the temporal dimension. For $E_D$ a second fully connected layer maps the pooled features of size 256 to 4 outputs, corresponding to the domain classification task. Similarly, $E_D$ transforms 256 pooled features to 2 outputs for PD task.
Training continued for 40 epochs; the batch size was set to 128.

\subsubsection{Evaluation metrics}
Most of the datasets for Parkinson's Disease detection are small. Therefore, we perform the stratified 5-fold cross-validation; each fold contains unique speakers to separate participants between train and test sets. For all metrics described below, we average metrics across folds. 
In order to evaluate the quality of predictions made by the models, we use several metrics: F1-score, Accuracy, Sensitivity, Positive Predictive Value (PPV), and Specificity. Accuracy is the percentage of correctly classified subjects. Specificity is the proportion of healthy subjects correctly classified, while sensitivity represents the percentage of PD patients correctly identified. Positive Predictive Value (PPV) is the proportion of correctly identified positive cases. F1 is the harmonic mean of PPV and sensitivity. Following \cite{ibarra2023towards}, we calculate all the metrics using per-person majority vote predictions (denoted $pp$ in tables) -- we aggregate all the model's predictions for each patient's utterances and assign the class that has the most predictions for that patient to obtain a single prediction -- whether a patient has PD or not.

\subsection{Results}

We compare the results of our model with baselines LSRC, Reddy \etal LSRC~\cite{Reddy}, Random Forst, SVM, and 1CNN, comparison results are presented in Table~\ref{tab:average_performance}. All the baselines were trained and evaluated in the same setting as our model -- 5-fold cross-validation with the same train-test split. We achieve state-of-the-art results in all tested metrics -- F1-score, Accuracy, Sensitivity, Positive Predictive Value, and Specificity, which show the strong capabilities of our approach and its generalizability to different languages.

\begin{table*}[!t]%
\centering %
\begin{minipage}{0.88\textwidth}
\caption{Overall performance comparison on PD-GITA, PD-Neurovoz, PD-Italian, MDVR-KCL datasets. Overall performance is computed as average across all datasets. \label{tab:average_performance}}%
\vspace{1.0em}
\begin{tabular*}{\textwidth}{@{\extracolsep\fill}llllll@{\extracolsep\fill}}
\toprule
\textbf{Method} & \textbf{F1pp $\uparrow$} & \textbf{Accpp $\uparrow$} & \textbf{SEpp $\uparrow$} & \textbf{PPVpp $\uparrow$} & \textbf{SPpp $\uparrow$} \\
\midrule
LSRC & 72.66 & 79.63 & 67.13 & 86.60 & 90.89 \\
Reddy \etal LSRC~\cite{Reddy} & 55.28 & 73.18 & 44.74 & 84.64 & 93.99 \\
Random Forest & 67.18 & 79.27 & 60.96 & 85.84 & \textbf{94.74} \\
SVM & 68.81 & 77.19 & 67.22 & 76.96 & 85.01 \\
1CNN~\cite{ibarra2023towards} & 80.78 & 79.11 & 75.84 & 79.95 & 83.07 \\
Ours DAPT+DAT & \textbf{89.28} & \textbf{92.00} & \textbf{91.22} & \textbf{90.55} & 92.13\\
\bottomrule
\end{tabular*}
\end{minipage}
\end{table*}

\subsubsection{Comprehensive Domain-Specific Performance Comparison}
Table~\ref{tab:performance} provides a detailed performance comparison of the different models across various domains. Each method's performance is evaluated on multiple datasets, including MDVR-KCL, PD-Italian, PD-Neurovoz, and PD-GITA, showing their effectiveness in PD detection in different languages and recording settings. The main set of metrics includes F1pp, Accpp, SEpp, PPVpp, and SPpp, representing per-person performance, where a majority vote over predictions for each patient was taken. Additional metrics include F1-score, Accuracy, Sensitivity (SE), Positive Predictive Value (PPV), and Specificity (SP); in this case, each metric was calculated in a per-audio-fragment manner with no aggregation across patients.

\begin{table*}[!t]
\small
\centering
\begin{minipage}{0.88\textwidth}
\caption{Comprehensive performance metrics for various methods and datasets, per-person metrics.\label{tab:performance}}
\vspace{1.0em}
\begin{tabular*}{\textwidth}{@{\extracolsep{\fill}} l l c c c c c}
\toprule
\textbf{Method} & \textbf{Dataset} & \textbf{F1pp $\uparrow$} & \textbf{Accpp $\uparrow$} & \textbf{SEpp $\uparrow$} & \textbf{PPVpp $\uparrow$} & \textbf{SPpp $\uparrow$} \\
\midrule
\multirow{5}{*}{LSRC} & MDVR-KCL & 63.63 & 78.11 & 47.70 & 100 & 100 \\
 & PD-Italian & 99.01 & 98.94 & 100 & 98.12 & 97.82 \\
 & PD-Neurovoz & 61.44 & 72.34 & 57.26 & 73.57 & 86.53 \\
 & PD-GITA & 66.57 & 69.13 & 63.57 & 74.70 & 79.19 \\
 & \textbf{Average} & \textbf{72.66} & \textbf{79.63} & \textbf{67.13} & \textbf{86.60} & \textbf{90.89} \\
\midrule
\multirow{5}{*}{Reddy \etal~\cite{Reddy}} & MDVR-KCL & 44.76 & 73.39 & 33.38 & 80.0 & 100 \\
 & PD-Italian & 90.14 & 89.38 & 83.29 & 100 & 100 \\
 & PD-Neurovoz & 45.01 & 70.86 & 31.74 & 81.28 & 96.78 \\
 & PD-GITA & 41.20 & 59.07 & 30.56 & 77.28 & 79.19 \\
 & \textbf{Average} & \textbf{55.28} & \textbf{73.18} & \textbf{44.74} & \textbf{84.64} & \textbf{93.99} \\
\midrule
\multirow{5}{*}{RF} & MDVR-KCL & 52.79 & 74.56 & 40.67 & 80.0 & 100 \\
 & PD-Italian & 97.05 & 96.78 & 96.43 & 98.12 & 97.82 \\
 & PD-Neurovoz & 44.79 & 69.97 & 32.77 & 80.04 & 95.18 \\
 & PD-GITA & 74.09 & 75.75 & 73.96 & 83.20 & 85.94 \\
 & \textbf{Average} & \textbf{67.18} & \textbf{79.27} & \textbf{60.96} & \textbf{85.84} & \textbf{94.74} \\
\midrule
\multirow{5}{*}{SVM} & MDVR-KCL & 63.09 & 76.37 & 58.75 & 80.92 & 94.51 \\
 & PD-Italian & 93.01 & 92.08 & 96.38 & 90.07 & 87.28 \\
 & PD-Neurovoz & 42.60 & 63.28 & 36.97 & 56.88 & 81.81 \\
 & PD-GITA & 76.52 & 77.01 & 76.78 & 77.99 & 76.44 \\
 & \textbf{Average} & \textbf{68.81} & \textbf{77.19} & \textbf{67.72} & \textbf{76.96} & \textbf{85.01} \\
\midrule
\multirow{5}{*}{1CNN~\cite{ibarra2023towards}} & MDVR-KCL & 81.14 & 82.54 & 81.67 & 50.00 & 83.00 \\
 & PD-Italian & 97.14 & 97.50 & 95.00 & 100.00 & 100.00 \\
 & PD-Neurovoz & 79.96 & 82.42 & 73.82 & 85.71 & 90.91 \\
 & PD-GITA & 64.88 & 53.97 & 52.86 & 84.08 & 58.35 \\
 & \textbf{Average} & \textbf{80.78} & \textbf{79.11} & \textbf{75.84} & \textbf{79.95} & \textbf{83.07} \\
\bottomrule
\end{tabular*}
\end{minipage}
\end{table*}

\subsubsection{Ablation Study}
Table~\ref{tab:ablation} shows the contribution of each part of the method to overall performance. All the experiments show average metrics across 4 datasets following the methodology described in the evaluation metrics section. First, we show that linear probing, i.e., using features directly from a frozen HuBERT encoder by feeding a classifier the features, gives a low performance. Then, we unfreeze the HuBERT weights and train the model as a whole, which marginally boosts performance. We show that DAPT should be used with attention to the dataset -- when pretraining on the mPower dataset, the quality of predictions drops -- which may be caused by two main reasons: lack of diversity in the dataset as it contains only iPhone recordings of people saying 'Aaaaah' and domain gap between mPower dataset and other PD datasets, where patients were saying words or sentences. As expected, when pretraining on AgeVox -- the domain gap between the dataset that HuBERT was initially trained on and PD datasets narrows -- and performance improves. Then, we demonstrate the importance of DAT -- which improves the accuracy of the model's predictions.

\begin{table*}[!t]%
\centering %
\begin{minipage}{0.88\textwidth}
\caption{Ablation Analysis. AgeVox DAPT represents domain adaptive Pretraining of HuBERT on AgeVox dataset, while mPower DAPT represents DAPT of HuBERT on mPower dataset. DAT represents further training on PD datasets with DAT. When DAT is not mentioned ~- training on PD datasets was done without DAT loss. Frozen HuBERT denotes the case when pretrained HuBERT encoder weights were frozen, and only the classification head was trained for PD detection task.\label{tab:ablation}}%
\vspace{1.0em}

\begin{tabular*}{\textwidth}{@{\extracolsep\fill}llllll@{\extracolsep\fill}}
\toprule
\textbf{Method} & \textbf{F1pp$\uparrow$} & \textbf{Accpp$\uparrow$} & \textbf{SE$\uparrow$} & \textbf{PPV$\uparrow$} & \textbf{SPpp$\uparrow$} \\
\midrule
AgeVox DAPT + DAT & \textbf{89.28} & \textbf{92.00} & \textbf{91.22} & 90.55 & \textbf{92.13} \\
DAT only & 88.15 & 91.77 & 86.13 & \textbf{92.41} & 93.64 \\
AgeVox DAPT only & 88.53 & 91.79 & 89.33 & 89.37 & 92.88 \\
no DAPT, no DAT & 84.21 & 89.56 & 85.47 & 89.92 & 90.58 \\
mPower DAPT + DAT & 81.43 & 87.22 & 80.79 & 84.02 & 89.55 \\
Frozen HuBERT & 66.72 & 62.71 & 84.84 & 59.26 & 48.65 \\
\bottomrule
\end{tabular*}
\end{minipage}
\end{table*}

\section{Discussion}

In this paper, we demonstrated that task-adaptive pre-training combined with classic domain adaptation techniques can boost the generalizability of automatic PD diagnosis from speech data. The proposed approach achieved state-of-the-art performance in PD detection tasks.
Leveraging a powerful pre-trained encoder, i.e., HuBERT, that was further improved for the task enabled generalization across datasets in different languages and settings.  This performance highlights its potential as a complementary tool for clinicians, particularly in early diagnosis, where identifying subtle speech abnormalities may lead to timely therapeutic interventions.

Despite its superior performance, the proposed approach has a number of limitations and requires further clinical validation prior to routine use in patient care. First, there is a need for domain adaptation when introducing a new group of subjects or novel protocols, which may require adjustments for populations with comorbidities that could affect speech, such as stroke or cognitive impairment. Second, the model does not assess symptom severity, which is crucial for tracking disease progression and tailoring treatment strategies in PD. Unlike previous studies that analyzed phonation or articulation patterns~\cite{AriasVergara2017}, it remains unclear which specific speech features are identified as markers of PD by the model, which limits its interpretability for clinicians.

Future work should prioritize building clinically interpretable outputs, such as identifying speech biomarkers linked to disease pathology or progression, to increase the tool's utility in practice. Ideally, an AI-based approach should complement a clinician’s assessment, offering objective, scalable evaluations that augment traditional diagnostic methods.

The labels that were used for model evaluation were mostly based on the Unified Parkinson’s Disease Rating Scale (UPDRS), which is known to have moderate inter-rater reliability~\cite{richards_interrater_1994}, potentially introducing variability into diagnostic criteria. Expanding datasets to include additional diagnostic information, such as genetic testing, neuroimaging, or patient-reported outcomes, could improve the reliability of the model’s predictions.

\section{Conclusion}
In this work, we presented a novel method combining Domain Adaptive Pretraining (DAPT) and Domain Adversarial Training (DAT) for a more generalizable detection of Parkinson's disease from raw speech data. We use HuBERT, a rich pre-trained speech encoder, and adapt it to speech from the elderly through domain adaptive pretraining on the Age-VOX-Celeb and mPower datasets. Our approach enables learning generalizable and language-independent deep features via DAT, improving the model's generalizability to unseen languages. Extensive evaluations on PD-GITA, Neurovoz, Italian, and MDVR-KCL datasets show the effectiveness of our method in detecting Parkinson's Disease.

\section*{Acknowledgments}
The work was partially sponsored by the Army Research Office and was accomplished under Cooperative Agreement Number W911NF-20-2-0053. The views and conclusions contained in this document are those of the authors and should not be interpreted as representing the official policies, either expressed or implied, of the Army Research Office or the U.S. Government. The U.S. Government is authorized to reproduce and distribute reprints for Government purposes, notwithstanding any copyright notation herein. We thank USC High-Performance Computing (HPC) for assisting with computational resources.

\section*{Declarations}

\begin{itemize}
\item \textbf{Funding}
Soleymani and Siniukov's Work was partially supported by The Army Research Office. Xing was supported by USC Viterbi CURVE fellowship. 
\item Competing interests 
The authors declare no conflict of interest.
\item Ethics approval and consent to participate
The work was a secondary analysis of anonymized publicly available data. 

\item \textbf{Data availability} 
The data is available from the original authors. The subset of VoxCeleb can be reproduced using the instructions available in the paper. 
\item \textbf{Code availability} 
Code and model weights will be made available on GitHub upon publication.
\item \textbf{Author contribution}
 Siniukov designed the machine learning approach, performed the experiments, and drafted the article. Xing re-implemented and conducted experiments for the baseline methods. Attaripour Isfahani contributed to drafting the paper and provided clinical input to the discussions. Soleymani has formulated the research and contributed to drafting the article.
\end{itemize}

\noindent

\bigskip

\bibliographystyle{unsrt} 
\bibliography{article-pd}

\begin{thebibliography}{10}

\bibitem{standaert2015parkinson}
David~G Standaert, Marie~H{\'e}l{\`e}ne Saint-Hilaire, and Cathi~A Thomas.
\newblock Parkinson’s disease handbook.
\newblock {\em American Parkinson Disease Association, New York, USA}, 2015.

\bibitem{birkmayer19613}
Walther Birkmayer and Oleh Hornykiewicz.
\newblock The l-3, 4-dioxyphenylalanine (dopa)-effect in parkinson-akinesia.
\newblock {\em Wiener klinische Wochenschrift}, 73:787--788, 1961.

\bibitem{kakkar2015management}
Ashish~Kumar Kakkar and Neha Dahiya.
\newblock Management of parkinson's disease: Current and future pharmacotherapy.
\newblock {\em European Journal of Pharmacology}, 750:74--81, 2015.

\bibitem{bosboom2003role}
JLW Bosboom, D~Stoffers, and E~Ch Wolters.
\newblock The role of acetylcholine and dopamine in dementia and psychosis in parkinson’s disease.
\newblock {\em Advances in Research on Neurodegeneration: Volume 10}, pages 185--195, 2003.

\bibitem{buddhala2015dopaminergic}
Chandana Buddhala, Susan~K Loftin, Brandon~M Kuley, Nigel~J Cairns, Meghan~C Campbell, Joel~S Perlmutter, and Paul~T Kotzbauer.
\newblock Dopaminergic, serotonergic, and noradrenergic deficits in parkinson disease.
\newblock {\em Annals of clinical and translational neurology}, 2(10):949--959, 2015.

\bibitem{delaville2011noradrenaline}
Claire Delaville, Philippe~de Deurwaerd{\`e}re, and Abdelhamid Benazzouz.
\newblock Noradrenaline and parkinson's disease.
\newblock {\em Frontiers in systems neuroscience}, 5:31, 2011.

\bibitem{fornai2007noradrenaline}
Francesco Fornai, Adolfo~B di~Poggio, Antonio Pellegrini, Stefano Ruggieri, and Antonio Paparelli.
\newblock Noradrenaline in parkinson's disease: from disease progression to current therapeutics.
\newblock {\em Current medicinal chemistry}, 14(22):2330--2334, 2007.

\bibitem{braga2019automatic}
Diogo Braga, Ana~M Madureira, Luis Coelho, and Reuel Ajith.
\newblock Automatic detection of parkinson’s disease based on acoustic analysis of speech.
\newblock {\em Engineering Applications of Artificial Intelligence}, 77:148--158, 2019.

\bibitem{coughlin2023fluid}
David~G Coughlin and David~J Irwin.
\newblock Fluid and biopsy based biomarkers in parkinson's disease.
\newblock {\em Neurotherapeutics}, 20(4):932--954, 2023.

\bibitem{donadio2020skin}
Vincenzo Donadio, Alex Incensi, Giovanni Rizzo, Rosa De~Micco, Alessandro Tessitore, Grazia Devigili, Francesca Del~Sorbo, Salvatore Bonvegna, Rossella Infante, Martina Magnani, Corrado Zenesini, Luca Vignatelli, Roberto Cilia, Roberto Eleopra, Gioacchino Tedeschi, and Rocco Liguori.
\newblock Skin biopsy may help to distinguish multiple system atrophy–parkinsonism from parkinson's disease with orthostatic hypotension.
\newblock {\em Movement Disorders}, 35(9):1649--1657, 2020.

\bibitem{little2008suitability}
Max Little, Patrick McSharry, Eric Hunter, Jennifer Spielman, and Lorraine Ramig.
\newblock Suitability of dysphonia measurements for telemonitoring of parkinson’s disease.
\newblock {\em Nature Precedings}, pages 1--1, 2008.

\bibitem{rusz2015speech}
Jan Rusz, Cecilia Bonnet, Jiří Klempíř, Tereza Tykalová, Eva Baborová, Michal Novotný, Aaron Rulseh, and Evžen Růžička.
\newblock Speech disorders reflect differing pathophysiology in {Parkinson}’s disease, progressive supranuclear palsy and multiple system atrophy.
\newblock {\em Journal of Neurology}, 262(4):992--1001, April 2015.

\bibitem{ibarra2023towards}
E.J. Ibarra, J.D. Arias-Londoño, M.~Zañartu, and J.I. Godino-Llorente.
\newblock Towards a corpus (and language)-independent screening of parkinson’s disease from voice and speech through domain adaptation.
\newblock {\em Bioengineering}, 10:1316, 2023.

\bibitem{Reddy}
M.~K. Reddy and P.~Alku.
\newblock Exemplar-based sparse representations for detection of parkinson's disease from speech.
\newblock {\em IEEE/ACM Transactions on Audio, Speech, and Language Processing}, 31:1386--1396, 2023.

\bibitem{sakar2013collection}
Betul~Erdogdu Sakar, M~Erdem Isenkul, C~Okan Sakar, Ahmet Sertbas, Fikret Gurgen, Sakir Delil, Hulya Apaydin, and Olcay Kursun.
\newblock Collection and analysis of a parkinson speech dataset with multiple types of sound recordings.
\newblock {\em IEEE Journal of Biomedical and Health Informatics}, 17(4):828--834, 2013.

\bibitem{berardelli2001pathophysiology}
Alfredo Berardelli, John~C Rothwell, Philip~D Thompson, and Mark Hallett.
\newblock Pathophysiology of bradykinesia in parkinson's disease.
\newblock {\em Brain}, 124(11):2131--2146, 2001.

\bibitem{jankovic2005motor}
Joseph Jankovic.
\newblock Motor fluctuations and dyskinesias in parkinson's disease: clinical manifestations.
\newblock {\em Movement disorders: official journal of the Movement Disorder Society}, 20(S11):S11--S16, 2005.

\bibitem{shahed2007motor}
Joohi Shahed and Joseph Jankovic.
\newblock Motor symptoms in parkinson's disease.
\newblock {\em Handbook of clinical neurology}, 83:329--342, 2007.

\bibitem{Amato2023}
F.~Amato, G.~Saggio, V.~Cesarini, G.~Olmo, and G.~Costantini.
\newblock Machine learning- and statistical-based voice analysis of parkinson’s disease patients: A survey.
\newblock {\em Expert Systems with Applications}, 219:119651, Jun. 2023.

\bibitem{Spanish}
Laureano Moro-Velazquez, Jorge~Andres Gomez-Garcia, Juan~Ignacio Godino-Llorente, Jesús Villalba, Jan Rusz, Stephanie Shattuck-Hufnagel, and Najim Dehak.
\newblock A forced gaussians based methodology for the differential evaluation of parkinson's disease by means of speech processing.
\newblock {\em Biomedical Signal Processing and Control}, 48:205--220, 2019.

\bibitem{ozbolt2022things}
Alex~S Ozbolt, Laureano Moro-Velazquez, Ioan Lina, Ankur~A Butala, and Najim Dehak.
\newblock Things to consider when automatically detecting parkinson’s disease using the phonation of sustained vowels: analysis of methodological issues.
\newblock {\em Applied Sciences}, 12(3):991, 2022.

\bibitem{Italian}
Giovanni Dimauro and Francesco Girardi.
\newblock Italian parkinson's voice and speech.
\newblock IEEE Dataport, Jun 2019.

\bibitem{ItalianPaper}
G.~Dimauro, V.~Di~Nicola, V.~Bevilacqua, D.~Caivano, and F.~Girardi.
\newblock Assessment of speech intelligibility in parkinson’s disease using a speech-to-text system.
\newblock {\em IEEE Access}, 5:22199--22208, 2017.

\bibitem{orozco2014new}
Juan~Rafael Orozco-Arroyave, Juli{\'a}n~David Arias-Londo{\~n}o, Jes{\'u}s~Francisco Vargas-Bonilla, Mar{\'\i}a~Claudia Gonzalez-R{\'a}tiva, and Elmar N{\"o}th.
\newblock New spanish speech corpus database for the analysis of people suffering from parkinson's disease.
\newblock In {\em LREC}, pages 342--347. ACL, 2014.

\bibitem{London}
Hagen Jaeger, Dhaval Trivedi, and Michael Stadtschnitzer.
\newblock Mobile device voice recordings at king's college london (mdvr-kcl) from both early and advanced parkinson's disease patients and healthy controls.
\newblock Zenodo, May 2019.

\bibitem{bot2016mpower}
Brian Bot, Christine Suver, Elias Chaibub~Neto, Michael Kellen, Arno Klein, J~Christopher Bare, Megan Doerr, Abhishek Pratap, John Wilbanks, E.~Dorsey, Stephen Friend, and Andrew Trister.
\newblock The mpower study, parkinson disease mobile data collected using researchkit.
\newblock {\em Scientific Data}, 3:160011, 03 2016.

\bibitem{quan2022end}
Changqin Quan, Kang Ren, Zhiwei Luo, Zhonglue Chen, and Yun Ling.
\newblock End-to-end deep learning approach for parkinson’s disease detection from speech signals.
\newblock {\em Biocybernetics and Biomedical Engineering}, 42(2):556--574, 2022.

\bibitem{karan2020hilbert}
Biswajit Karan, Sitanshu~Sekhar Sahu, Juan~Rafael Orozco-Arroyave, and Kartik Mahto.
\newblock Hilbert spectrum analysis for automatic detection and evaluation of parkinson’s speech.
\newblock {\em Biomedical Signal Processing and Control}, 61:102050, 2020.

\bibitem{karan2021improved}
Biswajit Karan and Sitanshu~Sekhar Sahu.
\newblock An improved framework for parkinson’s disease prediction using variational mode decomposition-hilbert spectrum of speech signal.
\newblock {\em Biocybernetics and Biomedical Engineering}, 41(2):717--732, 2021.

\bibitem{hawi2022automatic}
Sara Hawi, Jana Alhozami, Raneem AlQahtani, Dannah AlSafran, Maram Alqarni, and Lola El~Sahmarany.
\newblock Automatic parkinson’s disease detection based on the combination of long-term acoustic features and mel frequency cepstral coefficients (mfcc).
\newblock {\em Biomedical Signal Processing and Control}, 78:104013, 2022.

\bibitem{lahmiri2018performance}
Salim Lahmiri, Debra~Ann Dawson, and Amir Shmuel.
\newblock Performance of machine learning methods in diagnosing parkinson’s disease based on dysphonia measures.
\newblock {\em Biomedical engineering letters}, 8:29--39, 2018.

\bibitem{shahbakhi2014speech_geneticsvm}
Mohammad Shahbakhi, Danial~Taheri Far, and Ehsan Tahami.
\newblock Speech analysis for diagnosis of parkinson’s disease using genetic algorithm and support vector machine.
\newblock {\em Journal of biomedical science and engineering}, 2014, 2014.

\bibitem{nonavinakere2021detection}
Narendra Np, Björn Schuller, and Paavo Alku.
\newblock The detection of parkinson's disease from speech using voice source information.
\newblock {\em IEEE/ACM Transactions on Audio, Speech, and Language Processing}, PP:1--1, 05 2021.

\bibitem{tsanas2012novel}
Athanasios Tsanas, Max~A Little, Patrick~E McSharry, Jennifer Spielman, and Lorraine~O Ramig.
\newblock Novel speech signal processing algorithms for high-accuracy classification of parkinson's disease.
\newblock {\em IEEE transactions on biomedical engineering}, 59(5):1264--1271, 2012.

\bibitem{zhang2021parkinson}
Tao Zhang, Yajuan Zhang, Hao Sun, and Haoran Shan.
\newblock Parkinson disease detection using energy direction features based on emd from voice signal.
\newblock {\em Biocybernetics and Biomedical Engineering}, 41(1):127--141, 2021.

\bibitem{tuncer2020automated}
Turker Tuncer, Sengul Dogan, and Udyavara~Rajendra Acharya.
\newblock Automated detection of parkinson's disease using minimum average maximum tree and singular value decomposition method with vowels.
\newblock {\em Biocybernetics and Biomedical Engineering}, 40(1):211--220, 2020.

\bibitem{Velazquez}
L.~Moro-Velazquez, J.~A. Gomez-Garcia, J.~D. Arias-Londoño, N.~Dehak, and J.~I. Godino-Llorente.
\newblock Advances in parkinson’s disease detection and assessment using voice and speech: A review of the articulatory and phonatory aspects.
\newblock {\em Biomed. Signal Process. Control}, 66, 2021.

\bibitem{Little}
M.~A. Little, P.~E. McSharry, E.~J. Hunter, J.~Spielman, and L.~O. Ramig.
\newblock Suitability of dysphonia measurements for telemonitoring of parkinson’s disease.
\newblock {\em IEEE Trans. Biomed. Eng.}, 56(4):1015--1022, Apr 2009.

\bibitem{fang2020parkinsonian}
Hao Fang, Chen Gong, Chen Zhang, Yanan Sui, and Luming Li.
\newblock Parkinsonian chinese speech analysis towards automatic classification of parkinson's disease.
\newblock In {\em Machine Learning for Health}, pages 114--125. PMLR, 2020.

\bibitem{quan2021deep}
Changqin Quan, Kang Ren, and Zhiwei Luo.
\newblock A deep learning based method for parkinson’s disease detection using dynamic features of speech.
\newblock {\em IEEE Access}, 9:10239--10252, 2021.

\bibitem{er2021parkinson}
Mehmet~Bilal Er, Esme Isik, and Ibrahim Isik.
\newblock Parkinson’s detection based on combined cnn and lstm using enhanced speech signals with variational mode decomposition.
\newblock {\em Biomedical Signal Processing and Control}, 70:103006, 2021.

\bibitem{vasquez2018multitask}
Juan~Camilo V{\'a}squez-Correa, Tomas Arias-Vergara, Juan~Rafael Orozco-Arroyave, and Elmar N{\"o}th.
\newblock A multitask learning approach to assess the dysarthria severity in patients with parkinson's disease.
\newblock In {\em INTERSPEECH}, pages 456--460. ISCA, 2018.

\bibitem{klempivr2023evaluating}
Ond{\v{r}}ej Klemp{\'\i}{\v{r}}, David P{\v{r}}{\'\i}hoda, and Radim Krupi{\v{c}}ka.
\newblock Evaluating the performance of wav2vec embedding for parkinson's disease detection.
\newblock {\em Measurement Science Review}, 23(6):260--267, 2023.

\bibitem{klempir2024analyzing}
Ondrej Klempir and Radim Krupicka.
\newblock Analyzing wav2vec embedding in parkinson's disease speech: A study on cross-database classification and regression tasks.
\newblock {\em medRxiv}, pages 2024--04, 2024.

\bibitem{escobar2023deep}
Daniel Escobar-Grisales, Cristian~David R{\'\i}os-Urrego, and Juan~Rafael Orozco-Arroyave.
\newblock Deep learning and artificial intelligence applied to model speech and language in parkinson’s disease.
\newblock {\em Diagnostics}, 13(13):2163, 2023.

\bibitem{gunduz2019deep}
Hakan Gunduz.
\newblock Deep learning-based parkinson’s disease classification using vocal feature sets.
\newblock {\em IEEE Access}, 7:115540--115551, 2019.

\bibitem{wodzinski2019deep}
Marek Wodzinski, Andrzej Skalski, Daria Hemmerling, Juan~Rafael Orozco-Arroyave, and Elmar N{\"o}th.
\newblock Deep learning approach to parkinson’s disease detection using voice recordings and convolutional neural network dedicated to image classification.
\newblock In {\em 2019 41st annual international conference of the IEEE engineering in medicine and biology society (EMBC)}, pages 717--720. IEEE, 2019.

\bibitem{arias2020predicting}
Juli{\'a}n~D Arias-Londo{\~n}o and Jorge~A G{\'o}mez-Garc{\'\i}a.
\newblock Predicting updrs scores in parkinson’s disease using voice signals: A deep learning/transfer-learning-based approach.
\newblock In {\em Automatic Assessment of Parkinsonian Speech: First Workshop, AAPS 2019, Cambridge, Massachussets, USA, September 20--21, 2019, Revised Selected Papers 1}, pages 100--123. Springer, 2020.

\bibitem{sainath2015learning}
Tara~N Sainath, Ron~J Weiss, Andrew~W Senior, Kevin~W Wilson, and Oriol Vinyals.
\newblock Learning the speech front-end with raw waveform cldnns.
\newblock In {\em Interspeech}, pages 1--5. Dresden, Germany, 2015.

\bibitem{parisi2018feature}
Luca Parisi, Narrendar RaviChandran, and Marianne~Lyne Manaog.
\newblock Feature-driven machine learning to improve early diagnosis of parkinson's disease.
\newblock {\em Expert Systems with Applications}, 110:182--190, 2018.

\bibitem{vasquez2019convolutional}
Juan~Camilo V{\'a}squez-Correa, Tomas Arias-Vergara, Cristian~D Rios-Urrego, Maria Schuster, Jan Rusz, Juan~Rafael Orozco-Arroyave, and Elmar N{\"o}th.
\newblock Convolutional neural networks and a transfer learning strategy to classify parkinson’s disease from speech in three different languages.
\newblock In {\em Progress in Pattern Recognition, Image Analysis, Computer Vision, and Applications: 24th Iberoamerican Congress, CIARP 2019, Havana, Cuba, October 28-31, 2019, Proceedings 24}, pages 697--706. Springer, 2019.

\bibitem{rios2020transfer}
Cristian~David Rios-Urrego, Juan~Camilo V{\'a}squez-Correa, Juan~Rafael Orozco-Arroyave, and Elmar N{\"o}th.
\newblock Transfer learning to detect parkinson’s disease from speech in different languages using convolutional neural networks with layer freezing.
\newblock In {\em International Conference on Text, Speech, and Dialogue}, pages 331--339. Springer, 2020.

\bibitem{vasquez2021transfer}
Juan~Camilo V{\'a}squez-Correa, Cristian~David Rios-Urrego, Tomas Arias-Vergara, Maria Schuster, Jan Rusz, Elmar Noeth, and Juan~Rafael Orozco-Arroyave.
\newblock Transfer learning helps to improve the accuracy to classify patients with different speech disorders in different languages.
\newblock {\em Pattern Recognition Letters}, 150:272--279, 2021.

\bibitem{AriasVergara2017}
T.~Arias-Vergara, J.~C. Vásquez-Correa, and J.~R. Orozco-Arroyave.
\newblock Parkinson’s disease and aging: Analysis of their effect in phonation and articulation of speech.
\newblock {\em Cognitive Computation}, 9(6):731--748, Dec. 2017.

\bibitem{Garcia2021}
Adolfo~M. García, Tomás Arias-Vergara, Juan C.~Vasquez-Correa, Elmar Nöth, Maria Schuster, Ariane~E. Welch, Yamile Bocanegra, Ana Baena, and Juan~R. Orozco-Arroyave.
\newblock Cognitive determinants of dysarthria in parkinson's disease: An automated machine learning approach.
\newblock {\em Movement Disorders}, 36(12):2862--2873, 2021.

\bibitem{hubert}
Wei-Ning Hsu, Benjamin Bolte, Yao-Hung~Hubert Tsai, Kushal Lakhotia, Ruslan Salakhutdinov, and Abdelrahman Mohamed.
\newblock Hubert: Self-supervised speech representation learning by masked prediction of hidden units.
\newblock {\em IEEE/ACM Transactions on Audio, Speech, and Language Processing}, 29:3451--3460, 2021.

\bibitem{dontstoppretrain}
Suchin Gururangan, Ana Marasovi{\'c}, Swabha Swayamdipta, Kyle Lo, Iz~Beltagy, Doug Downey, and Noah~A Smith.
\newblock Don't stop pretraining: Adapt language models to domains and tasks.
\newblock {\em arXiv preprint arXiv:2004.10964}, 2020.

\bibitem{ganin2015unsupervised}
Yaroslav Ganin and Victor Lempitsky.
\newblock Unsupervised domain adaptation by backpropagation.
\newblock In {\em International conference on machine learning}, pages 1180--1189. PMLR, 2015.

\bibitem{voxceleb}
Arsha Nagrani, Joon~Son Chung, Weidi Xie, and Andrew Zisserman.
\newblock Voxceleb: Large-scale speaker verification in the wild.
\newblock {\em Computer Speech \& Language}, 60:101027, 2020.

\bibitem{chung2018voxceleb2}
Joon~Son Chung, Arsha Nagrani, and Andrew Zisserman.
\newblock Voxceleb2: Deep speaker recognition.
\newblock {\em arXiv preprint arXiv:1806.05622}, 2018.

\bibitem{tawara2021agevox}
Naohiro Tawara, Atsunori Ogawa, Yuki Kitagishi, and Hosana Kamiyama.
\newblock Age-vox-celeb: Multi-modal corpus for facial and speech estimation.
\newblock In {\em ICASSP 2021-2021 IEEE International Conference on Acoustics, Speech and Signal Processing (ICASSP)}, pages 6963--6967. IEEE, 2021.

\bibitem{mfcc_davis1980comparison}
Steven Davis and Paul Mermelstein.
\newblock Comparison of parametric representations for monosyllabic word recognition in continuously spoken sentences.
\newblock {\em IEEE transactions on acoustics, speech, and signal processing}, 28(4):357--366, 1980.

\bibitem{panayotov2015librispeech}
Vassil Panayotov, Guoguo Chen, Daniel Povey, and Sanjeev Khudanpur.
\newblock Librispeech: an asr corpus based on public domain audio books.
\newblock In {\em 2015 IEEE international conference on acoustics, speech and signal processing (ICASSP)}, pages 5206--5210. IEEE, 2015.

\bibitem{kahn2020libri}
J.~Kahn, M.~Riviere, W.~Zheng, E.~Kharitonov, Q.~Xu, P.E. Mazare, J.~Karadayi, V.~Liptchinsky, R.~Collobert, C.~Fuegen, T.~Likhomanenko, G.~Synnaeve, A.~Joulin, A.~Mohamed, and E.~Dupoux.
\newblock Libri-light: A benchmark for asr with limited or no supervision.
\newblock In {\em ICASSP 2020-2020 IEEE International Conference on Acoustics, Speech and Signal Processing (ICASSP)}, pages 7669--7673. IEEE, 2020.

\bibitem{willis2013parkinson}
Allison~W Willis.
\newblock Parkinson disease in the elderly adult.
\newblock {\em Missouri medicine}, 110(5):406, 2013.

\bibitem{richards_interrater_1994}
Marcus Richards, Karen Marder, Lucien Cote, and Richard Mayeux.
\newblock Interrater reliability of the unified {Parkinson}'s disease rating scale motor examination.
\newblock {\em Movement Disorders}, 9(1):89--91, 1994.
\newblock \_eprint: https://onlinelibrary.wiley.com/doi/pdf/10.1002/mds.870090114.

\end{thebibliography}

\end{document}